\documentclass[fleqn,usenatbib]{mnras}

\usepackage{newtxtext,newtxmath}

\usepackage[T1]{fontenc}

\DeclareRobustCommand{\VAN}[3]{#2}
\let\VANthebibliography\thebibliography
\def\thebibliography{\DeclareRobustCommand{\VAN}[3]{##3}\VANthebibliography}

\usepackage{graphicx}	%
\usepackage{amsmath}	%

\usepackage{pdflscape}
\usepackage{booktabs}

\usepackage{tablefootnote}
\usepackage{soul}
\usepackage{mathtools}
\newcounter{tableeqn}[table]

\newcounter{tablesubeqn}[tableeqn]

\usepackage{tikz}
 
\usepackage{multirow}

\title[Helium triplet observability of young planets]{Helium escape signatures are generally strongest during younger ages but this age dependence is lost in the diversity of observed exoplanets  }%

\author[A. P. Allan \& A. A. Vidotto]{
Andrew P. Allan,$^{1}$\thanks{E-mail: allan@strw.leidenuniv.nl} 
Aline A. Vidotto,$^{1}$
\\
$^{1}$Leiden Observatory, Leiden University, P.O. Box 9513, 2300 RA Leiden,
The Netherlands\\ }

\date{Accepted 2025 March 31. Received 2025 March 7; in original form 2025 January 10}

\pubyear{2025}

\begin{document}
\label{firstpage}
\pagerange{\pageref{firstpage}--\pageref{lastpage}}
\maketitle

\begin{abstract}
Highly irradiated exoplanets undergo extreme hydrodynamic atmospheric escape, due to their high level of received XUV flux. Over their lifetime, this escape varies significantly, making evolution studies essential for interpreting the growing number of observations of escaping planetary atmospheres. In a previous work, we modelled this evolving escape, alongside one of its observable tracers, the helium triplet transit signature at 1083nm. Using hydrodynamic and ray-tracing models, we demonstrated that atmospheric escape and the corresponding He 1083nm signature are stronger at younger ages, for a 0.3$~M_\text{J}$ gas-giant. Yet, the current literature includes several young (<1Gyr) planets with weak or non-detections in He 1083nm. To understand this apparent discrepancy, we now perform detailed modelling for many of these systems. The resulting He 1083nm predictions align relatively well with the observations. 
From our two studies, we conclude that for any given planet, stronger atmospheric escape during younger ages produces deeper He 1083nm absorption. However, for a population of exoplanets, the relation between younger ages and stronger He absorptions is lost to the broad diversity of their various other system parameters. %
Accordingly, for the current sample of young, 1083nm-observed exoplanets, alternative trends take precedence. One such trend is that planets with deeper geometrical transits exhibit more favourable detections.
Our modelling also agrees with the strong empirical trend in the literature between $ EW \cdot R_{*}^{2}$ and $F_{\text{xuv}} \cdot R_{\text{pl}}^2 / \Phi_{g}$. 
Additionally, we show that the coupling between the lower and upper atmospheres is necessary for a robust prediction of the 1083nm signature. %
\end{abstract}


\section{Introduction} 
\label{sec:intro}

Exoplanets close in to their host star receive a high flux of X-ray and ultraviolet (XUV) radiation capable of heating their planetary atmosphere and ultimately driving its hydrodynamic escape. Numerous confirmed detections of escaping atmospheres have been made using the method of transit spectroscopy. This was first achieved via the hydrogen Lyman-$\alpha$ line for the hot-Jupiter HD$~$209458b \citep{Vidal_Madjar2003_first_detect_escaping_H_HD209}. The helium triplet at 1083$~$nm \citep{Seager_Sasselov_2000, Oklopcic_2018_10839_window} has since overtaken Lyman-$\alpha$ as the most popular tracer for escaping atmospheres \citep[for review, see][]{Dos_Santos_2022_iauga_obs_of_pl_winds_outflows}. 
This near-infrared spectral feature is produced by helium in the metastable 2$^3$S state absorbing stellar photons resulting in the 2$^3$S to 2$^3$P transition. 
This observing strategy is still in its infancy, with the first space-based detection only happening in 2018 \citep{Spake_2018_NATURE_He_in_atm}, swiftly followed by ground-based detections \citep{Nortmann_2018, Allart_2018_warm_Nep_Hat_p_11b_He_obs}. Despite its novelty, there have now been over 50 helium triplet transit observations due to its ability to be observed using ground-based telescopes. Hence, population studies linking planetary parameters with the observed helium triplet absorption are now feasible \citep{Krishnamurthy_Cowan_2024}. Such studies benefit from homogeneous surveys obtained with the same instruments \citep{Zhang_4_mini_Nep, Allart_2023_SPIRou_He_11_gas_giants, Vissapragada_2022_survey, Orell-Miquel_2024_MOPYS}.

Unfortunately, observing escaping atmospheres with the helium triplet entails many difficulties, as proven by the currently greater number of non-detections than detections. The high uncertainty of the XUV flux received by the planet, a parameter which greatly affects both the escape and its observability, is one of the most difficult problems currently being faced. The ability of stellar variability to convincingly mimic the desired signature of planetary atmospheric escape is another. On the latter however, \citet{krolikowski2024strength} show that the intrinsic stellar variability should not preclude detection of young exospheres, except at the youngest ages when the star is most active. The current scarcity of suitable close-in, young, transiting systems as well as poor constraints on their planetary masses also adds to the difficulty of observing and modelling atmospheric escape at young ages.

Despite the mentioned observational difficulties, younger ages (< 1$~$Gyr) remain a highly interesting phase in the context of planetary atmospheric escape, given that this is when the escape is strongest. At these ages, the high XUV flux and weaker gravitational force due to the `puffed up' nature of younger exoplanets leads to a stronger atmospheric escape \citep{Allan_Vidotto_2019}. With time, the escape weakens due mainly to the reduction of XUV flux received as the star spins down \citep{Vidotto_2014} and the shrinking of the planetary radius as the planet cools \citep{Fortney2010}. Hence, the first Gyr of evolution is when the most drastic atmospheric changes occur \citep{Owen_Wu_2013_Kepler_pl_tale_of_evol}. The extent to which a planetary atmosphere escapes depends on many factors, with the planetary gravity, the level of XUV flux received and the reservoir of atmospheric material all playing pivotal roles. In the most extreme cases, the primordial atmosphere can be lost entirely, changing the very nature of the planet \citep{Lopez_2012_thermal_evol_mass_loss_sculpt_pop, Owen_Wu2013, Fossati_2017, Daria_2018_grid_1_40_earths}. Due to such cases, the process of extreme hydrodynamic escape is one of the leading theories currently offered to explain the sub-Jovian desert and radius gap \citep{Fulton_2017} apparent in the current population of observed exoplanets \citep{ Owen_Wu_2017_evap_valley_kelper, Jin_and_Mordasini_2018}. %

Given the importance of atmospheric escape at younger ages, studies are beginning to purposefully target young exoplanets for helium triplet observations. \citet{Zhang_4_mini_Nep} reported escaping helium in four <1$~$Gyr mini-Neptunes using Keck/NIRSPEC. \citet{Alam_2024_young_non_detects} set upper-limits due to non-detections in three <800$~$Myr sub-Jovian exoplanets with Keck II/NIRSPEC. Gathering published helium observations, \citet{Krishnamurthy_Cowan_2024} show that large planets have detections across young and old ages while small (<2.6 R$_{\oplus}$) planets only have detections at young ages. They attribute this to the total loss of the planetary atmosphere for smaller planets. The MOPYS (Measuring Out-flows in Planets orbiting Young Stars) project of \citet{Orell-Miquel_2024_MOPYS} performed high-resolution spectroscopy observations with CARMENES \citep{Quirrenbach_2014CARMENES_instrument_overview} and GIARPS \citep{GIARPS_2016} for 20 exoplanets. Their large scale project found that 0.1-1$~$Gyr-old planets do not exhibit more helium triplet and hydrogen H-$\alpha$ detections than older planets, in agreement with the study of \citet{Krishnamurthy_Cowan_2024}. This absence of strong trend between planetary age and atmospheric escape signature is interesting considering that numerous theoretical studies escape predict escape to be strongest at the youngest ages. For example, the evolution studies of \citet{Allan_Vidotto_2019} and \citet{Allan_et_al_He_evol_2024} show clear trends between younger ages and stronger hydrogen (Lyman-$\alpha$ and H-$\alpha$) and helium triplet absorption. 

In this current work, we seek to clear up this apparent discrepancy. 
We do so by running self-consistent hydrodynamic escape models for numerous young planets which have helium triplet observations. In section \ref{sec:modelling}, we outline our modelling approach. In section \ref{sec: results discussion}, we present our findings, discussing trends between planetary parameters and both the modelled and observed triplet absorption. Finally, we summarise our conclusions in section \ref{sec:conclusions}.

\section{Modelling approach}

\label{sec:modelling}

 We utilise two models in this work. The first being our hydrodynamic model, which self-consistently solves the fluid dynamic equations in addition to the coupled equations for hydrogen ionisation balance and helium state populations. It solves for transitions between the helium $1^1S$, $2^1S$, $2^3S$ and singly and doubly ionised states. The second is our ray-tracing model which utilises the atmospheric profiles and calculates the individual extinction by each of the lines of the helium triplet during a planetary transit. We refer the reader to \citet{Allan_et_al_He_evol_2024} for more in-depth descriptions of both models. The ray-tracing technique remains unchanged. In this current work, we outline only additions to our hydrodynamic model. 
The main additions can be summarised as follows. We now: 
\begin{itemize}
\item model a sample of observed young systems (section \ref{sec:planet_sample}), rather than the lifetime evolution of fictitious highly irradiated exoplanets \citep{Allan_Vidotto_2019,Allan_et_al_He_evol_2024}.

\item make a more physically-informed, system-dependant predictions for the conditions at the lower boundary of our modelled atmospheres (section \ref{sec:determine_atm_base_low_BC}). 

\item extend our modelled atmospheres to the minimum height required to fully encompass the stellar disk rather than $10R_\text{pl}$, given the smaller size of these modelled planets. 

\end{itemize}

\subsection{Sample of modelled systems}
\label{sec:planet_sample}

As previously discussed in section \ref{sec:intro}, young exoplanets are of particular interest when it comes to understanding the helium triplet transit signature both observationally and theoretically. Hence, we choose to model younger planets, with our chosen sample satisfying the following requirements: 
\begin{itemize}
\item An age estimate below 1 Gyr\footnote{We make an exception for TOI-2018b, given that it was initially presented as a young planet, although now has an updated, relatively young age estimate of 1.6 to 2.6 Gyr. TOI-1683b is a similar case, with \citet{Zhang_4_mini_Nep} quoting a young age of 500 $\pm$ 150 Myr while \citet{Orell-Miquel_2024_MOPYS} recently pushed this back to 2$^{+1.3}_{-0.9}~$Gyr.  }. 
\item An orbital distance below 0.15 au.
\item A host star of spectral-type K or G. 
\item A high-resolution helium triplet observation.
\end{itemize}

\begin{table*}
\caption{Properties of the planets considered in this work. Each planets has \
helium triplet transmission spectroscopy observations. The columns refer to planetary name, classification, age, mass, radius, orbital distance, \
and transit impact parameter and transit duration from first to fourth contact. \
Planetary mass values marked with superscripts `a' and `b' indicate that the mass was obtained via the mass-radius relation of \citet{Chen_Kipping_2017_mass_radius_relation} and \citet{Wolfgang_2016_mass_radius_relation_sub_nep}, respectively, \
due to lack of a precise radial velocity measurement.}
\label{Tab:pl_inputs}
\begin{tabular}{lllllllll}
\toprule
name & classification & age & $M_{ \text{pl} }$ & $R_{\text{pl} }$ & a & b & $t_{\rm dur}$ & Reference \\ 
 &  & (Myr) & ($M_{\oplus} )$ & ($R_{\oplus} )$ & (AU) &  & (hours) &  \\ \midrule
Wasp-52b & inflated hot-Saturn & $400^{+300}_{-200}$ & 137.9 & 14.0 & 0.027 & 0.60 & 1.81 & \citep{Kirk_2022_Wasp52b_Wasp177b} \\
V1298$~$Tau$~$c & proto-sub-Neptune & 23$\pm$4 & 26.7$^{\text{a}}$ & 5.5 & 0.082 & 0.35 & 4.66 & \citep{Alam_2024_young_non_detects} \\
TOI-1268b & hot-Saturn & 110-380 & 96.3 & 9.1 & 0.072 & 0.42 & 4.00 & \citep{Perez_Gonzales_2023_TOI-1268b} \\
TOI-1683b & mini-Neptune & 500$\pm$150 & 8.0$^{\text{b}}$ & 2.3 & 0.036 & 0.85 & 1.28 & \citep{Zhang_4_mini_Nep} \\
K2-100b & Neptune desert dweller & 750$\pm$5 & 21.8 & 3.9 & 0.030 & 0.79 & 1.61 & \citep{Alam_2024_young_non_detects} \\
TOI-1430b & mini-Neptune & 165$\pm$30 & 7.0$^{\text{b}}$ & 2.1 & 0.070 & 0.44 & 2.71 & \citep{Zhang_4_mini_Nep} \\
TOI-2018b & super-Earth & 2400$^{+200}_{-800}$ & 9.2 & 2.3 & 0.061 & 0.55 & 2.36 & \citep{2023AJ....166...49D} \\
TOI-2076b & mini-Neptune & 204$\pm$50 & 9.0$^{\text{b}}$ & 2.6 & 0.063 & 0.34 & 3.25 & \citep{Zhang_4_mini_Nep} \\
HD$~$63433b & mini-Neptune & 414$\pm$23 & 5.3$^{\text{a}}$ & 2.2 & 0.072 & 0.38 & 2.93 & \citep{Alam_2024_young_non_detects} \\
HD$~$63433c & mini-Neptune & 414$\pm$23 & 7.3$^{\text{a}}$ & 2.7 & 0.146 & 0.37 & 4.10 & \citep{Zhang_2022_HD63433_upperlimit_non_detect} \\
TOI-560b & mini-Neptune & 480-750 & 15.9 & 2.8 & 0.060 & 0.57 & 2.14 & \citep{Zhang_4_mini_Nep} \\
K2-136c & sub-Neptune & 650$\pm$70 & 18.1 & 3.0 & 0.110 & 0.31 & 3.45 & \citep{Mayo_2023_K2136} \\
\bottomrule \end{tabular}
\end{table*}

\begin{table*}
\caption{Properties of the stars used in this work. The columns give the stellar spectral-type, age, mass, radius and the luminosity in each of the XUV and mid-UV bins (specified in the text). The first three host stars listed have their own SED from the literature. For the remainder, we utilise proxy SEDs from stars with similar spectral type and radius as indicated by the final column. MUSCLES series refers to the following works: \citet{France_2016, Youngblood_2016_MUSCLES_CITE2, Loyd_2016_muscles_cite3}. MUSCLES extension refers to \citet{Behr_2023_muscles_extension}.}
\label{Tab:stellar_SED}
\begin{tabular}{llllllllll}
\toprule
star & type & age & $M_{ \text{*} }$  & $R_{\text{*} }$ & $L_{\text{X-ray}}$ & $L_{\text{hEUV}}$ & $L_{\text{sEUV}}$ & $L_{\text{mid-UV}}$ & SED reference \\ 
 &  & (Myr) & ($M_{\odot})$ & ($R_{\odot})$ & ($10^{-5}$L$_{\odot})$ & ($10^{-5}$L$_{\odot})$ & ($10^{-5}$L$_{\odot})$ & ($10^{-5}$L$_{\odot})$ &  \\ \midrule
  \multicolumn{6}{c}{\textbf{Stars with an XUV SED available}} \\ \
V1298 Tau & G2 & 23$\pm$4 & 1.10 & 1.34 & 91.16 & 18.62 & 8.59 & 591 & \citep{Duvvuri_2023_high_energy_sepc_of_v1298_tau} \\
HD 63433 & G5 & 414$\pm$23 & 0.99 & 0.91 & 2.04 & 1.75 & 2.55 & 2017 & \citep{Zhang_2022_HD63433_upperlimit_non_detect} \\
TOI-560 & K4V & 480-750 & 0.73 & 0.65 & 0.35 & 0.42 & 0.44 & 289 & \citep{Zhang_2022_TOI_560} \\
 \multicolumn{6}{c}{\textbf{Stars requiring a proxy star}} \\ \
K2-100 & G0 & 750$\pm$5 & 1.15 & 1.24 & 1.31 & 1.13 & 3.05 & 664 & $\iota$-Horologii \\
TOI-1268 & K1/K2 & 110-380 & 0.96 & 0.92 & 0.72 & 1.05 & 1.20 & 517 & $\epsilon$-Eridani \\
Wasp-52 & K2 & $400^{+300}_{-200}$ & 0.87 & 0.79 & 0.53 & 0.78 & 0.88 & 382 & $\epsilon$-Eridani \\ 
TOI-1683 & K0 & 500$\pm$150 & 0.69 & 0.64 & 0.34 & 0.50 & 0.57 & 247 & $\epsilon$-Eridani \\
TOI-1430 & K2V & 165$\pm$30 & 0.85 & 0.78 & 0.52 & 0.76 & 0.87 & 376 & $\epsilon$-Eridani \\
TOI-2076 & K0 & 204$\pm$50 & 0.82 & 0.76 & 0.49 & 0.72 & 0.82 & 355 & $\epsilon$-Eridani \\
TOI-2018 & KV & 2400$^{+200}_{-800}$ & 0.57 & 0.62 & 0.02 & 0.12 & 0.10 & 30 & HAT-P-12 \\
K2-136 & K5.5 & 650$\pm$70 & 0.74 & 0.68 & 0.03 & 0.14 & 0.12 & 35 & HAT-P-12 \\
 \multicolumn{6}{c}{\textbf{Proxy stars}} \\ \
$\iota$-Horologii\tablefootnote{We used the same SED as \citet{Alam_2024_young_non_detects}, in which the \citet{SForcada2019} model was joined with the solar SED for the longer wavelengths.} & G0V & 600 & 1.34 & 1.13 & 1.31 & 1.13 & 3.05 & 664 & \citep{SForcada2019} \\
$\epsilon$-Eridani & K2 & 600$\pm$200 & 0.82 & 0.74 & 0.47 & 0.68 & 0.77 & 335 & MUSCLES series:[V2.2] \\
HAT-P-12 & K5 & 2500$\pm$2000 & 0.73 & 0.70 & 0.03 & 0.15 & 0.13 & 38 & MUSCLES extension:[V2.4] \\
\bottomrule \end{tabular}
\end{table*}

\begin{figure}
    \includegraphics[width=0.45\textwidth]{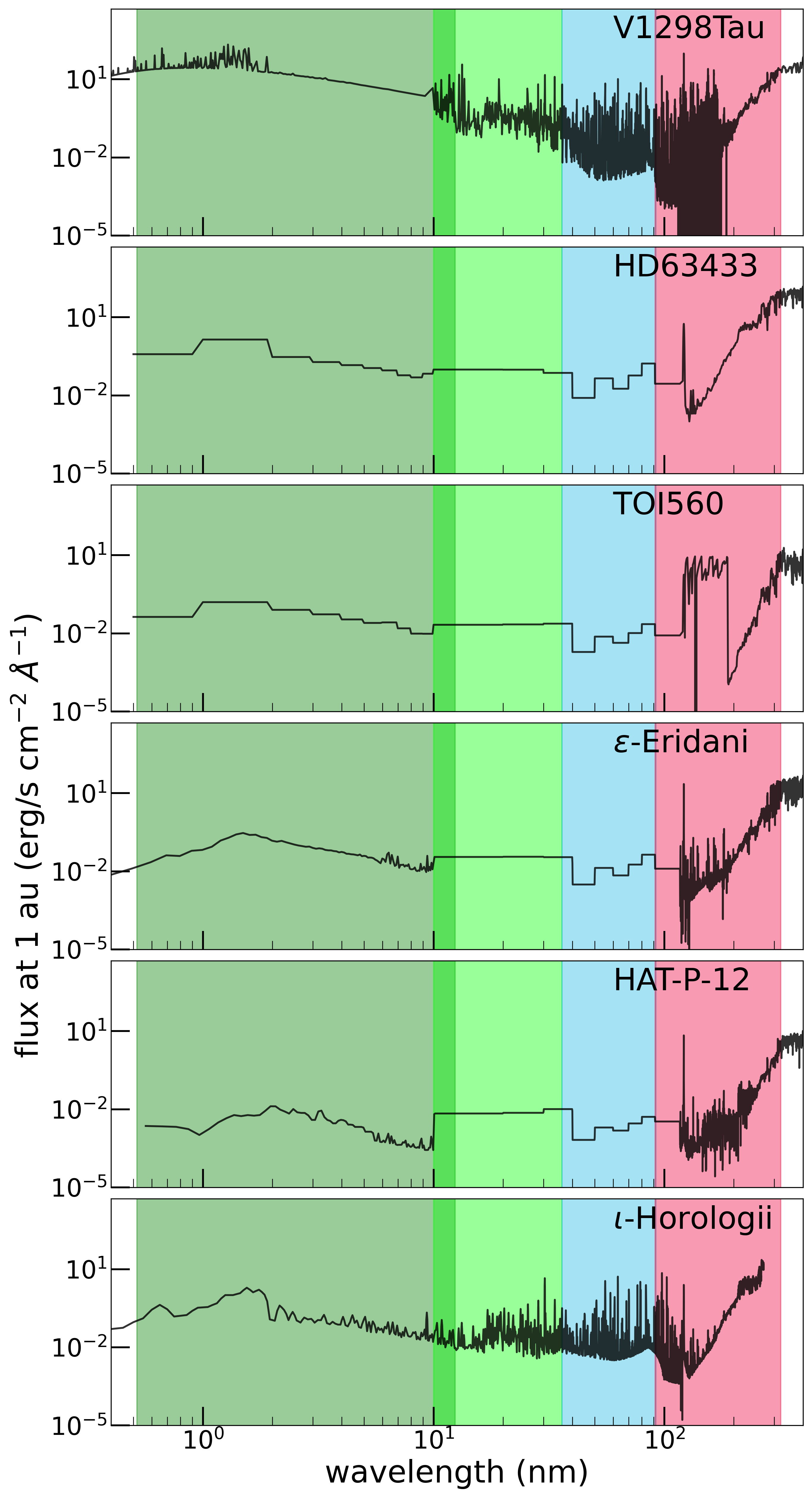}
    \caption{Stellar spectral energy distributions (SED's) adopted in our modelling. Table \ref{Tab:stellar_SED} lists the stellar parameters of each as well as the relevant reference. The shaded regions from left to right denotes the X-ray, hard-EUV, soft-EUV and mid-UV wavelength bins.} %
    \label{fig:SEDs}
\end{figure}

This selection process led us to 12 planets in our sample. Table \ref{Tab:pl_inputs} lists all of these planets along with their planetary parameters required as input to our modelling. Of all of the considered planets, only three of their host stars have a spectral energy distribution (hereby SED) published in the literature. For the remaining stars, we use their spectral type to assign them a proxy SED. We use the SED to calculate the stellar flux at the planet's orbital distance. Figure \ref{fig:SEDs} displays each of the SEDs utilised. From left to right, they are binned into X-ray (0.517-12.4$~$nm), hard-EUV (10-36$~$nm), soft-EUV (36-92$~$nm) and mid-UV (91.2-320$~$nm) in accordance with \citet{Allan_et_al_He_evol_2024} and \citet{Johnstone_2021}. The resulting luminosities are listed in Table \ref{Tab:stellar_SED} along with other assumed stellar properties relevant to our modelling.   

\subsection{Determining the hydrodynamic atmospheric base}
\label{sec:determine_atm_base_low_BC}

An important assumption of hydrodynamic escape models is the location of the upper atmospheric base, as well as its density and temperature. Usually, models are computed starting at the geometric radius of the planet, hereby $R_{\text{pl}}$. The sensitivity of the predicted hydrodynamic escape to boundary conditions assumed at the modelled upper atmospheric base (hereby $R_{\text{base}}$) has been studied by many. 
For example, the predicted atmospheric escape of the purely hydrogen, photons concentrated on a single energy, self-consistent, 1$~$D model of \citet{Murray-Clay2009} was shown to be reasonably insensitive to the assumed base density $\rho[R_{\text{base}}]$, base H-ionisation fraction $f_{H^+}[R_ \text{base}]$ and their base temperature $T[R_{\text{base}}]$, so long as; $\tau[R_{\text{base}}] >> 1$; $f_{H^+}[R_\text{base}]<< 1$ and $T[R_{\text{base}}] << 10^4~$K.
 They do, however, note their solution to be highly sensitive to the height designated to the upper atmosphere base, with a 10\% uncertainty in this producing as much as a factor of two variation in the predicted mass-loss rate. Using an order of magnitude calculation, they show the uncertainty in this height to be \textasciitilde10\%, given that their $R_{\text{base}}$ is set at $1R_{\text{pl}}$ rather 1.1$R_{\text{pl}}$, where they find UV photons are absorbed. Indeed, applying the method later outlined in this section to the HD$~$209458b analogue of \citet{Murray-Clay2009}, we reach only a slightly higher atmospheric base of $R_{\text{base}}=1.18~R_{\text{pl}}$. Given their determined uncertainty of 10\% in $R_{\text{base}}$, they determine their mass-loss rate predictions to include an uncertainty of at most a factor of two. For some of our considered sample of young exoplanets however, we calculate considerably larger relative values for $R_{\text{base}}$ (see Table \ref{table_lower_boundary_conditions}), raising the importance of the chosen height for the base of the upper atmosphere, due to its effect on the predicted atmospheric escape and resulting signatures.
\citet{Salz2016b} reported similar findings to \citet{Murray-Clay2009} using the Pluto-Cloudy Interface \citep[TPCI][]{Salz_2015_PLUTO_CLOUDY_Interface}, with sensitivities to $T[R_{\text{base}}]$ and $\rho[R_{\text{base}}]$ resulting in respective uncertainties of $<$10\% and $<$50\% in the predicted mass-loss rate.

We previously confirmed the discussed lower boundary sensitivities findings of \cite{Murray-Clay2009, Salz2016b} in our purely hydrogen, photons concentrated at a single energy, self-consistent, 1$~$D model \citep{Allan_Vidotto_2019}. When including helium species and splitting the flux into various energy bins, we again found general agreement with the mentioned sensitivities \citep[see appendix A of][]{Allan_et_al_He_evol_2024}. Interestingly however, we also noted our model to exhibit a strong sensitivity of the predicted helium triplet profile with the assumed boundary conditions, beyond the uncertainties of the observations (now demonstrated again in Appendix \ref{sec:appendix_old_BC} for models of TOI-1286b and TOI-560b). For example, raising the assumed density at the base of the modelled atmosphere by an order of magnitude was found to increase the equivalent width of the predicted helium triplet by factors of 2.7 and 1.8 at respective ages of 16$~$Myr and 5$~$Gyr \citep{Allan_et_al_He_evol_2024}. 
This sensitivity has important implications in the context of modelling the observable helium triplet signature, particularly of younger planets and motivated us to better constrain the assumed boundary conditions at the base of our modelled upper atmosphere. 

\begin{table}
\caption{Upper atmospheric base properties, where the base is set at a pressure of 10 nanobar, and the density and temperature  were calculated as explained in section \ref{sec:determine_atm_base_low_BC}. The final two columns give the base density assuming He/H number fractions of 0.1 and 0.02, respectively, with their difference arising from their mean molecular weights. \
              }
\label{table_lower_boundary_conditions}
\begin{tabular}{lllll}
\toprule
name & $R_{\text{base}}$ &  $T[R_{\text{base}}]$ &  $ \rho[R_{\text{base}}]$ &  $ \rho[R_{\text{base}}]$ \\ 
 & $(R_{\text{pl}}$) & (K) & (g cm$^{-3}$) & (g cm$^{-3}$) \\ \midrule
Wasp-52b & 1.25 & 987 & 1.60e-13 & 1.30e-13 \\
V1298$~$Tau$~$c & 1.43 & 829 & 1.90e-13 & 1.55e-13 \\
TOI-1268b & 1.16 & 676 & 2.33e-13 & 1.90e-13 \\
TOI-1683b & 1.51 & 666 & 2.37e-13 & 1.93e-13 \\
K2-100b & 1.65 & 1375 & 1.15e-13 & 9.35e-14 \\
TOI-1430b & 1.49 & 604 & 2.61e-13 & 2.13e-13 \\
TOI-2018b & 1.32 & 479 & 3.29e-13 & 2.68e-13 \\
TOI-2076b & 1.49 & 650 & 2.42e-13 & 1.98e-13 \\
HD$~$63433b & 1.77 & 729 & 2.16e-13 & 1.76e-13 \\
HD$~$63433c & 1.49 & 512 & 3.08e-13 & 2.51e-13 \\
TOI-560b & 1.26 & 541 & 2.91e-13 & 2.38e-13 \\
K2-136c & 1.18 & 402 & 3.92e-13 & 3.20e-13 \\
\bottomrule \end{tabular}
\end{table}

In order to achieve this, we utilise inputs from a lower atmosphere model. The lower and upper atmospheric models are coupled at a pressure $P[R_{\text{base}}]$ of 10 nanobar. We select this pressure to correspond to a height in the atmosphere at which heat from XUV photoionisation is sufficiently deposited. This chosen pressure was influenced by the few nanobar quoted in \citet[][sketch in figure 8]{Murray-Clay2009} for where UV photons are absorbed and \citet{Tang_fortney_murray_clay} who state that the photoionising high-energy flux is deposited at a nanobar pressure. 
We note however that other studies assume greatly differing pressures at the base of their modelled escaping atmospheres, such as \citet{Koskinen_2013} who select 1$~\mu$bar, based on photo-chemical calculations of HD$~$209458b \citep{Lavvas_2014_HD209458b}. %
\citet{Yan_2022_Wasp_52b} also assume a base pressure of 1$~\mu$bar, while \citet{Lampon_2020} instead use 1$~$mbar. %
The \textsc{ATES} photoionization hydrodynamics code \citep[ATmospheric EScape, ][]{Caldiroli_2021_ATES} follows the approach of \citet{Salz2016b} for the lower boundary conditions, assuming a total number density of $10^{14}$ cm$^{-3}$ (corresponding to pressures around 14$~\mu$bar) at a modelled atmospheric base set to 1$~R_{\text{pl}}$.
\citet{Salz2016b} highlight the slight inconsistency in beginning the simulation at 1$~R_{\text{pl}}$ despite the assumed density being reached at a height above the planetary surface. This is currently a common inconsistency, present also in the previous versions \citep{Allan_Vidotto_2019,Allan_et_al_He_evol_2024} of the model presented here. Improving upon this by now coupling lower and upper atmospheric models,
we utilise the lower atmosphere model of \citet{Parmentier_Guillot_I}. This analytical model provides the temperature-optical depth relation for planetary atmospheres that are heated both from below (intrinsic heat) and above (stellar irradiation), assuming a plane-parallel atmosphere inspired by Eddington's approximation. As input we provide the planetary gravity $g$ and the irradiated temperature, calculated using equation 1 of \citet{Guillot_2010}. For all remaining inputs, we default to the standard values given in the \citet{Parmentier_Guillot_I} model. 

From the static model, we obtain atmospheric profiles of pressure $P_\text{s}$, temperature $T_\text{s}$, mean molecular weight $\mu_\text{s}$, and optical depth $\tau_{\text{s}}$ where subscripts `s' indicates static rather than hydrodynamic output profiles. We refer to the atmospheric height where $\tau_{\text{s}}=2/3$ as the planetary photosphere, setting $R_{\text{photo}}=1~R_\text{pl}$.         
Under the assumptions of hydrostatic equilibrium for an isothermal gas, we then calculate the height $z$ above the photosphere at which our chosen $P[R_{\text{base}}]=10$ nanobar is achieved: 
        \begin{equation}    
z=-H_{\text{s}} \Big( \ln(P[R_{\text{base}}])  - \ln(P[\tau_{\text{s}}=2/3]) \Big),  
        \end{equation}
where $H_{\text{s}} =\frac{k~T_{\text{s}}}{\mu_{\text{s}} m_\text{H} g}$ is the atmospheric scale height, with $k$ being the Boltzmann constant and $\mu_{\text{s}}$ being the mean molecular weight in the lower static atmosphere.  
We set the base of our hydrodynamic atmosphere to occur at the height $R_{\text{base}}=1~R_{\text{pl}}+z$, corresponding to a pressure of 10 nanobar. This height and the corresponding temperature and density (from the ideal gas law) required by our modelling are listed in Table \ref{table_lower_boundary_conditions} for each planet. For our sample, $R_{\text{base}}$ spans from 1.16 to 1.77 $R_\text{pl}$, considerably above the standardly assumed $1~R_\text{pl}$ lower boundary condition used in our previous \citet{Allan_et_al_He_evol_2024} model version. While still inevitably making an assumption about the environment at the base of the atmosphere (namely the pressure being 10$~$nanobar), this approach allows for a more physically informed estimate, given that the planetary gravity and the irradiated temperature now inform our choice of both the pressure and temperature for our hydrodynamic model.

As input for our ray-tracing model for predicting the helium triplet transit signature, we fuse the atmospheric profiles of the static atmosphere (spanning $1~R_\text{pl}$ to $R_\text{base}$) with the hydrodynamic atmosphere. In doing so, we assume that the fraction of helium in the triplet state within the innermost static atmosphere is constant and equal to the value predicted by the hydrodynamic atmosphere at $R_{\text{base}}$. Testing our ray-tracing model by assuming various constants for this fraction in the static atmosphere, we determine this value to only negligibly affect the resulting helium triplet profile predictions.

A comparison between the predicted triplet profiles obtained using the commonly used boundary conditions of $R_\text{base}=1~R_\text{pl}$, $\rho[R_{\text{base}}]=4 \times 10^{-14}$g$~$cm$^{-3}$, $T[R_{\text{base}}]=1000~$K rather than those assumed in this work, is presented in section \ref{sec:appendix_old_BC} of the appendix.

\section{Results \& Discussion}
\label{sec: results discussion}

\subsection{General hydrodynamic predictions}

\label{sec:atmo_escape_predictions}

 \begin{table}
\caption{Atmospheric escape predictions for each of our planets. $\dot{m}$ refers to the mass-loss rate. We define $v_{\text{term}}$ as the velocity reached at 10$~R_{\text{pl}}$. The hydrodynamic atmospheric temperature profile starts at $T [R_{\rm base}]$, reaching a maximum temperature  max($T$) and cools off beyond that. EW refers to the equivalent width of the predicted helium triplet signature,
while peak refers to its peak excess absorption. Predictions are giving first assuming a He/H number fraction of 0.1 and then 0.02.}
\label{Tab:escape_predictions}
\begin{tabular}{llllll}
\toprule
name & $\dot{m}$ & $v_{\text{term}}$ & max($T$) & EW & peak \\ 
 & ($10^{12}~$g/s) & (km/s) & (kK) & (m\AA ) & (\%) \\ \midrule
\multicolumn{4}{c}{\textbf{He/H=0.1}} \\ 
Wasp-52b & 2.0 & 67 & 11.2 & 269 & 16.7 \\
V1298$~$Tau$~$c & 6.5 & 31 & 11.1 & 147 & 7.4 \\
TOI-1268b & 0.5 & 23 & 10.1 & 49 & 4.9 \\
TOI-1683b & 3.4 & 27 & 9.9 & 16 & 1.5 \\
K2-100b & 7.6 & 34 & 11.3 & 23 & 1.2 \\
TOI-1430b & 1.9 & 21 & 7.6 & 8.4 & 1.0 \\
TOI-2018b & 0.3 & 11 & 4.6 & 5.5 & 0.9 \\
TOI-2076b & 2.0 & 23 & 8.4 & 7.6 & 0.8 \\
HD$~$63433b & 5.7 & 26 & 10.3 & 7.5 & 0.8 \\
HD$~$63433c & 1.4 & 20 & 7.3 & 4.8 & 0.7 \\
TOI-560b & 0.6 & 19 & 7.0 & 5.0 & 0.7 \\
K2-136c & 0.06 & 8 & 5.2 & 2.8 & 0.6 \\
\multicolumn{4}{c}{\textbf{He/H=0.02}} \\ 
Wasp-52b & 2.6 & 69 & 11.3 & 113 & 8.8 \\
V1298$~$Tau$~$c & 6.2 & 34 & 11.2 & 48 & 2.5 \\
TOI-1268b & 0.5 & 28 & 10.1 & 12 & 1.4 \\
TOI-1683b & 4.1 & 29 & 9.9 & 6.2 & 0.8 \\
K2-100b & 7.1 & 37 & 11.3 & 5.8 & 0.4 \\
TOI-1430b & 2.3 & 21 & 7.7 & 2.6 & 0.4 \\
TOI-2018b & 0.4 & 12 & 4.5 & 1.7 & 0.3 \\
TOI-2076b & 2.5 & 23 & 8.6 & 2.7 & 0.4 \\
HD$~$63433b & 8.6 & 26 & 10.5 & 3.7 & 0.5 \\
HD$~$63433c & 2.0 & 20 & 7.6 & 2.0 & 0.4 \\
TOI-560b & 0.6 & 21 & 6.6 & 1.2 & 0.2 \\
K2-136c & 0.06 & 8 & 4.1 & 0.6 & 0.2 \\
\bottomrule \end{tabular}
\end{table}

Table \ref{Tab:escape_predictions} lists predicted properties for each of our modelled exoplanets; the mass-loss rate ($\dot{m}$), the wind terminal velocity ($v_{\text{term}}$), the peak of the temperature profile and the equivalent width and peak excess absorption of the helium triplet profiles. 
 The adopted helium fraction has little effect on the predicted escape properties while naturally having a strong correlation on the corresponding observable helium signature. The mass-loss rate predictions appear to be strong relative to some recent studies of the same planets \citep{Zhang_4_mini_Nep, McCreery_2025}. %
As discussed in more detail in section \ref{sec:compar_predict_mdot_McCreery} of the Appendix, differing modelling natures are likely responsible for this noted discrepancy.

Overall, we find that the hydrodynamic behaviour of the escaping atmospheres modelled here follows that described previously in \citet{Allan_et_al_He_evol_2024}, for an evolving, theoretical, 0.3$~$M$_{\text{Jup}}$ planet closely orbiting a K-type star. Accordingly, we now only briefly discuss the dominant hydrodynamic processes.

For our entire exoplanet sample, heating within the majority of the modelled atmosphere is predominantly driven by the photoionisation of neutral hydrogen H$^0$ by sEUV photons. Within some inner region specific to each planet but extending out to $\lesssim 2 ~ R_{\text{pl}}$ for all, alternative photoionisations dominate the heating. Assuming He/H=0.02, these are the photoionisations of H$^0$ by hEUV photons and by 24.6$~$eV photons (themselves emitted in direct recombinations to He(1$^1$S) from He$^+$). These photoionisations contribute comparable heating rates within this mentioned inner atmospheric region. Assuming instead He/H=0.1, the hEUV photoionisation of H$^0$ and He(1$^1$S) are the strongest heaters within this region. The described heating behaviour is consistent with that shown in \citet[][upper panels of figures 5 and B1]{Allan_et_al_He_evol_2024}. The non-negligible contribution of the 24.6$~$eV photon to atmospheric heating and consequent escape is of interest, given that it is often omitted from similar models. 

Atmospheric heating is partially countered by cooling, with adiabatic expansion being the most effective cooling mechanism for each of the modelled planets. Lyman-$\alpha$ cooling also contributes substantially for the hotter planets, Wasp-52b, V$~$1298$~$Tau and K2-100b. For these planets, Lyman-$\alpha$ cooling exceeds adiabatic cooling within a small inner region below where the outflow becomes supersonic, very similar to that shown for the younger (hotter) planet in \citet[][lower panel of figure 5]{Allan_et_al_He_evol_2024}. This noted role of Lyman-$\alpha$ cooling for hotter exoplanets is in agreement with the models of \citet{Murray-Clay2009} and \textsc{ATES} \citep{Caldiroli_2021_ATES}.

\subsection{ He triplet transmission spectroscopy predictions }
\label{sec:triplet_predictions}

To model helium triplet transit signatures, we first compute the population of helium in the 2$^3$S state. %
This again follows a similar behaviour as that described in \citet{Allan_et_al_He_evol_2024}. Namely, the recombination from ionised helium to the helium triplet state is the dominant populating pathway across all atmospheric heights. A major depopulating path is collisions with free electrons in to the He($2^1S$) state. Photoionisations out of the triplet state mostly by mid-UV photons also contribute non-negligibly to the depopulation of the 2$^3$S state, in the more tenuous outer regions of the atmospheres.

\begin{figure*}
    \includegraphics[width=0.95\textwidth]{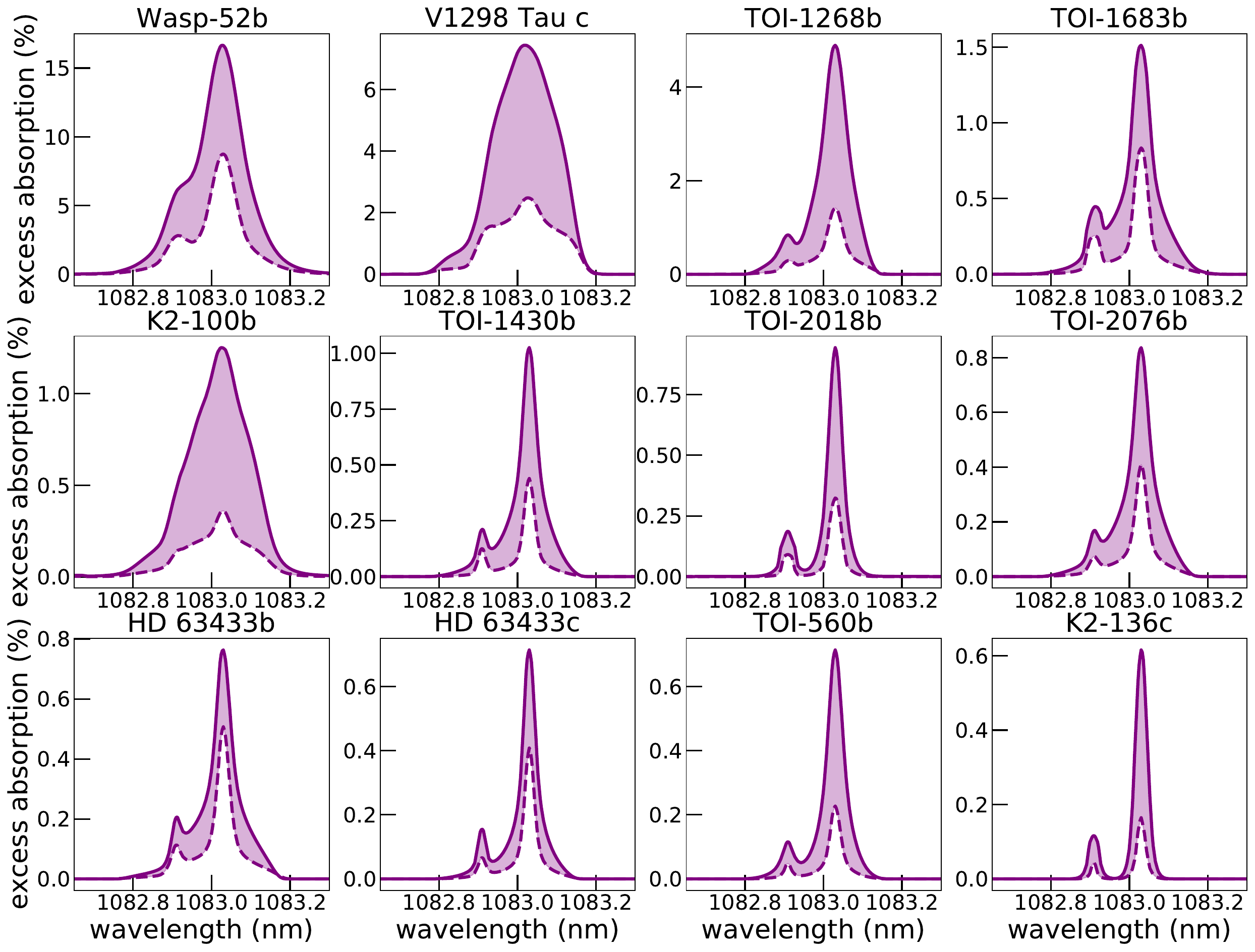}
    \caption{ Model predictions of the helium 1083$~$nm transmission spectra for each planet, ordered by declining predicted absorption. The solid and dashed lines show models assuming helium to hydrogen number fractions of He/H=0.1 and 0.02, respectively. The profiles are averaged over transit phases between first and fourth contacts. Note the strong variations between the various y-axes. }
 \label{fig:all_helium_profiles_10nanobar}
\end{figure*}

Figures \ref{fig:all_helium_profiles_10nanobar} displays our modelled helium triplet profiles for each considered planet. The solid and dashed lines were obtained by assuming constant number fractions of He/H=0.1 and 0.02 respectively. We average the predicted profiles for all phases between first and fourth contacts of the planetary transit, as done in \citet{Allan_et_al_He_evol_2024}, in order to account for transit observations being integrated over a time sufficient to obtain a high signal-to-noise, rather than instantaneously at mid-transit. It is clear that there is great diversity in the shapes of the predicted profiles. The helium triplet consists of three lines, at 1083.034$~$nm, 1083.025$~$nm, and 1082.909$~$nm \citep[wavelengths in air][]{NIST_ASD} where the first two are indistinguishable due to their proximity. It is interesting to note that for K2-100b, V1298$~$Tau$~$c, Wasp-52b all three lines blend together. This is the result of stronger Doppler broadening, with the atmospheric outflows of all three planets achieving fast velocities before the density falls beyond that required for the helium triplet profile to be affected. This is due to their relatively strong mass-loss rates combined with their fast terminal velocities, each above 30km/s (see Table \ref{Tab:escape_predictions}). It is also worth highlighting that the assumed He/H fraction more strongly affects some of the predicted triplet profiles, as is the case for TOI-560b and TOI-1268b. While for HD$~$63433b and c, it has a much weaker effect.  %

\begin{figure}
    \includegraphics[width=0.48\textwidth]{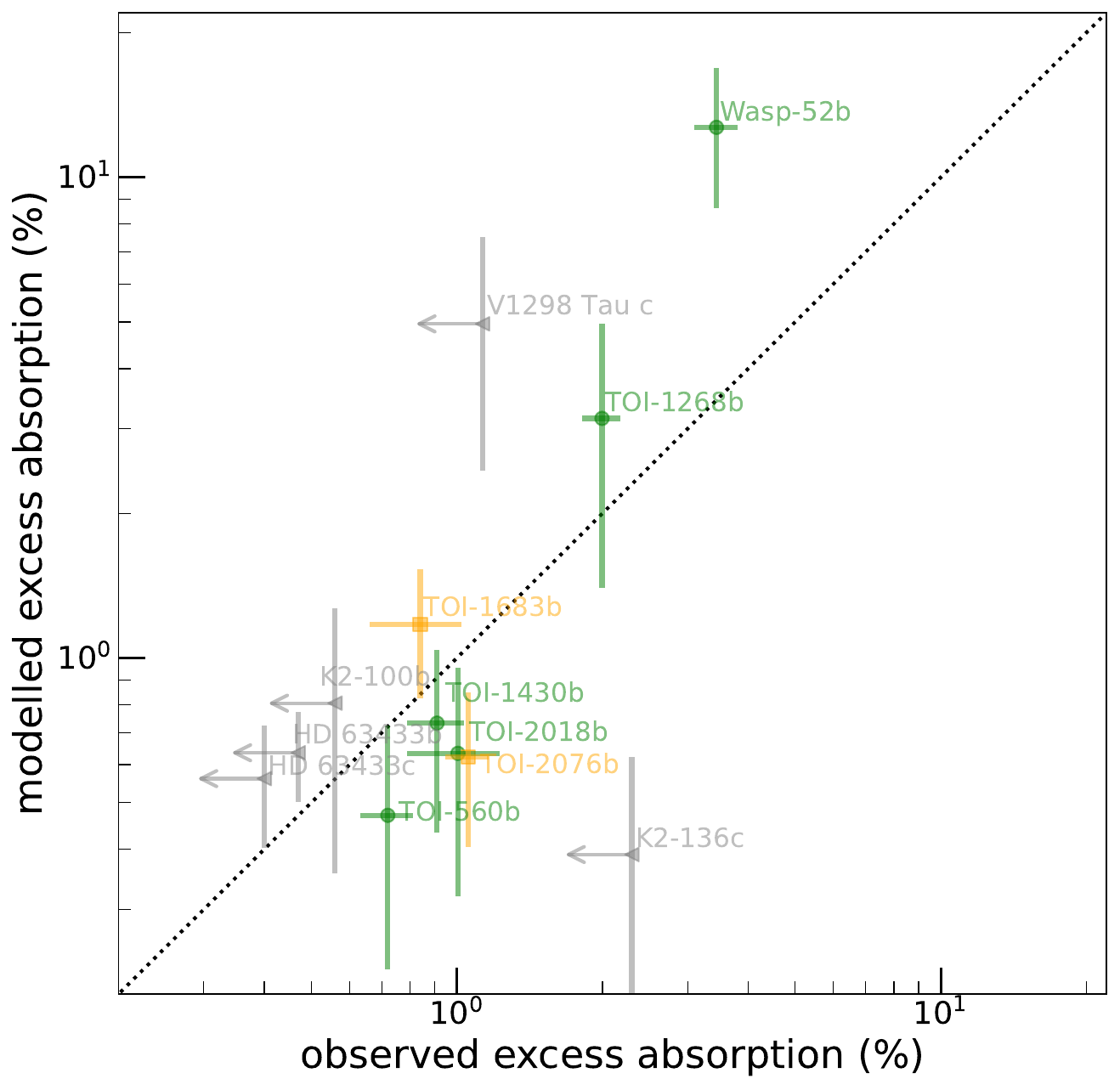}
    \caption{ The relation between the observationally detected helium triplet excess absorption (green), conflicting detection (yellow) or upper limit due to non-detection (grey with arrow) with the modelled-predicted excess absorption for each planet. The diagonal dashed line indicates a perfect 1:1 match of the model prediction to the observation. Table \ref{Tab:observational} gives more detail on each observation. For each planet, the spread in the predicted excess absorption corresponds to assuming a He/H number fraction ranging from 0.02 to 0.1. }
 \label{fig:modelled_vs_observed_absorption}
\end{figure}

We compare the model-predictions to the observed excess absorptions in figure \ref{fig:modelled_vs_observed_absorption}. In this and subsequent figures, we distinguish between planets with confirmed detections, those with conflicting detections (see Table \ref{Tab:observational}) and observational upper-limits due to non-detections, by the colours green, amber and grey. The vertical spread in the predicted excess absorption corresponds to assuming a He/H number fraction ranging from 0.02 to 0.1.
Overall, figure \ref{fig:modelled_vs_observed_absorption} shows a positive agreement between the observations and model predictions.

Despite the overall agreement, for some planets, V$~1298~$Tau$~$c in particular, there is disagreement between the model predicted and observed helium triplet profiles. While our model predicts strong 2 to 7\% absorption assuming He/H respective fractions of 0.02 to 0.1, \citet{Alam_2024_young_non_detects} set an upper limit of 1.1\% with Keck II/NIRSPEC. There are various potential difficulties and limitations associated with performing helium triplet transit signatures of young exoplanets, any one of which could be responsible for this noted discrepancy.
One particular difficulty is the presence of stellar activity, which is higher at such young ages. With an estimated age of 30$~$Myr, V$~1298~$Tau$~$c is the youngest planet in our sample and a high level of activity of V$~1298~$Tau$~$ has been previously shown by \citet{krolikowski2024strength} and \citet{2021_Vissapragada_search_He_in_V1298_tau_system} with, for example, flare detection. High stellar activity can produce significant variability in the chromospheric stellar helium triplet, which can both mask or mimic the helium triplet signature of escaping planetary material \citep{krolikowski2024strength}. It can also introduce variability in the planetary escape signature itself by altering the level of XUV flux received. High stellar activity also complicates determining the planetary mass by affecting the radial velocity signatures. Accordingly, the literature currently offers no known planetary mass for the planets of V$~1298~$Tau, as is the case for five of the other young planets we model (see Table \ref{Tab:pl_inputs}). In such cases, studies modelling the atmospheric escape and corresponding signatures must resort to mass-radius relations \citep
 {Chen_Kipping_2017_mass_radius_relation} in order to obtain a planetary mass estimate. However, such mass-radius relations are better suited to more mature planets. As \citet{Poppenhaeger_2021} show, the uncertainty in the planetary masses of the V$~1298~$Tau system can have drastic effects on each planets mass-loss rate and overall evolution. As a consequence, this could also affect the predicted helium transit signatures. 
 
 In addition to the mentioned observational difficulties associated with young ages, various limitations within the model can also cause discrepancy between the observed and predicted helium signatures. These limitations stem mostly from the hydrodynamic models
1$-$D nature, forcing the omission of various relevant physical processes. Star-planet interactions involving the stellar wind and planetary magnetic fields require 3$-$D magneto-hydrodynamic modelling. These interactions have been shown capable of strongly affecting the Lyman-$\alpha$ \citep{Carolan_2021_plB_field, Presa_2024_atm_esc_HJ_subalfven} and helium triplet signatures \citep{Schreyer_2024_helium_10830_constrain_pl_B_fields} of planetary atmospheric escape. 
An interesting point related to this is that the three planets in our sample with the largest model-predicted mass-loss rates, K2-100b, HD$~$63433b and V$~$1298$~$Tau$~$c, have tightly constrained non-detections for the helium triplet. We speculate that their non-detections could be due to stellar winds. It is known that stellar wind mass-loss rates increase with stellar X-ray fluxes \citep{Wood2002_mdot_fucntion_age_activity, vidotto2021_review} and that stellar winds can confine the extension of planetary atmospheres reducing the amount of absorption seen in spectroscopic transits \citep{Cleary_vidotto2020, Carolan_2021_SW}. It is therefore suggestive that V$~$1298$~$Tau, K2-100\footnote{for which $\iota$-Horologii was used as a proxy (see table \ref{Tab:stellar_SED}).} and HD$~$63433, with their largest X-ray surface fluxes in our sample, could also generate strong stellar winds that would confine (or reduce the extension) of the helium-absorbing atmosphere.
While beyond the scope of this work, the mentioned potential effects of stellar wind interactions on the observables is worth noting, particularly considering that the magnetic fields of both low-mass stars and gaseous exoplanets are expected to be strongest during their younger ages \citep{Skumanich_1972_spin_down_of_stars, Vidotto_2014_empirical_trends_age_rotation, Kilmetis_2024_B_field_evol}. 
Another possible problem in the model is in estimating the total helium abundance, which could in reality differ from that of our assumed constant He/H number fractions  of 0.02 and 0.1. Finally, it is very important to have a robust characterisation of the stellar high-energy spectrum, due to its importance not only in driving the escape hydrodynamically, but also affecting the photoionisation of helium in the triplet state. Unfortunately, the stellar high-energy spectrum is rather difficult to obtain.

We highlight an interesting difference in the behaviour of the modelled hot-Saturns, Wasp-52b and TOI-1268b, and the remaining smaller planets. In order for our model predicted triplet profiles of these two giants to reproduce their observed values, a low He/H number fraction of $\lesssim$ 2\% is required. Whereas $\gtrsim$ 10\% He/H is required for the remaining smaller planets. 
Planetary formation theory suggests that primordial planetary atmospheres accrete hydrogen and helium from the protoplanetary disk with He/H number fraction close to the solar value of $\sim0.1$ \citep{PhysRevD.92.123526}. However, fits of various model to observed helium triplet data in the literature suggest sub-solar He/H for the escaping atmospheres of many exoplanets, including HD$~189733$b and GJ$~3470$b \citep{Lampon_2021}. Such estimates have generally been made using single-fluid modelling with a constant He/H fraction assumed throughout the escaping atmosphere, as is the case for our model. \citet{Xing_mass_fractionation_2023} however have performed multi-fluid modelling in order to demonstrate for HD$~$209458b, the escaping atmosphere can have sub-solar He/H even with an approximately solar helium and hydrogen abundances, due to mass-fractionation in the upper escaping atmosphere. Hence, fits suggesting low He/H does not necessarily mean sub-solar fractions of hydrogen and helium accreted during planetary formation. The preferential loss of hydrogen has also been shown capable of strongly enhancing the He/H fraction of planets on the upper edge of the radius valley \citep{Malsky_2023_Helium_enhanced}. These planets are susceptible to this given the loss of significant fractions of their initial atmosphere as mentioned in section \ref{sec:intro}. While considerably younger than the planets of the modelled grid of \citet{Malsky_2023_Helium_enhanced}, our models apparent preference of higher He/H for smaller mini-Neptune relative to that for the hot-Saturns, could be explained by a similar enhancement. Given their relatively small reservoirs of atmospheric material and their strong predicted mass-loss rates, the enhancement of their He/H fraction to above that of the larger hot-Saturns may occur sufficiently quick to be noticeable even at such young ages. While a preferential loss of hydrogen may also occur for larger hot-Saturn size planets, their large  atmospheric reservoirs would likely prevent significant enhancement of He/H. In order to confirm this physical interpretation, both the expansion of the planet sample and evolutionary atmospheric escape modelling using a multi-fluid approach would be valuable.

\subsection{Relations with the helium triplet signature}
\label{sec: trends}

\begin{figure*}
    \includegraphics[width=1.0\textwidth]{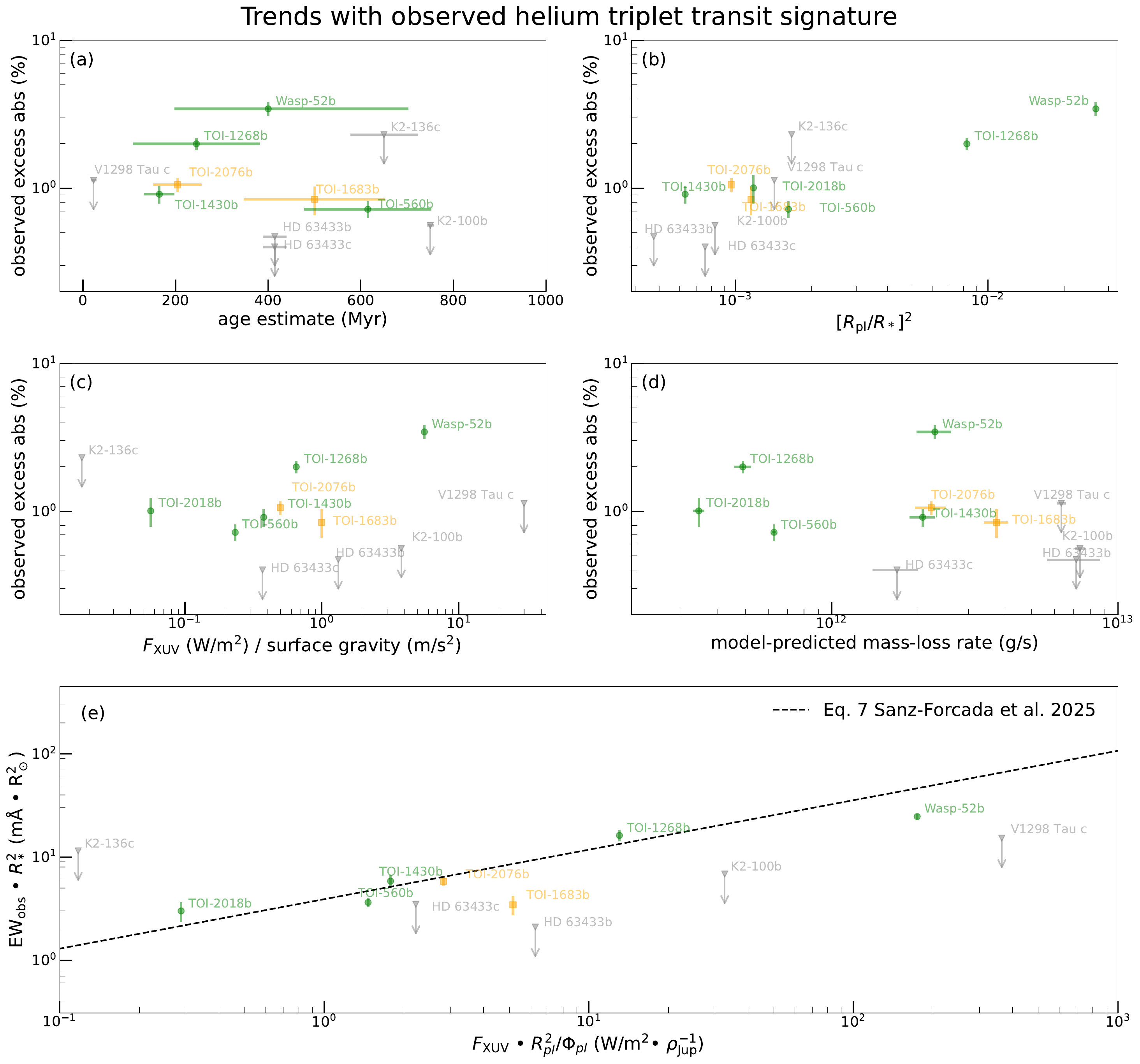}
    \caption{ The relation between various system parameters denoted by the x-axes with observational parameters of the helium triplet transit signature on the y-axes. Planets with successful detections are marked by green circles, those with conflicting interpretations (see Table \ref{Tab:observational}) are marked by amber squares while upper-limits due to non-detections are indicated by grey triangles. The y-axes of panels (a-d) display the observed excess absorption while panel (e) displays the observed equivalent width of the triplet multiplied by the stellar disk area $R_*^2$, with the dashed line displaying the trend of \citet[][their equation 7]{Sanz-Forcada_2025_HeI_10830}. The spread in the predicted mass-loss rates corresponds to a range of He/H number fractions of 0.02 to 0.1, whereas the spread in age, observed excess absorption and equivalent width is simply that reported by their references and given in tables \ref{Tab:pl_inputs} and \ref{Tab:observational}. For clarity, TOI-2018b is omitted from panel (a) due to its older age. Similarly, the mass-loss rate of K2-136c (which observationally produced a non-detection) is below the displayed range of mass-loss rates in panel (d). Tables \ref{Tab:pl_inputs}, \ref{Tab:escape_predictions}, \ref{Tab:observational} contain their omitted values. }
 \label{fig:trends_obs}
\end{figure*}

\begin{figure*}
    \includegraphics[width=1.0\textwidth]{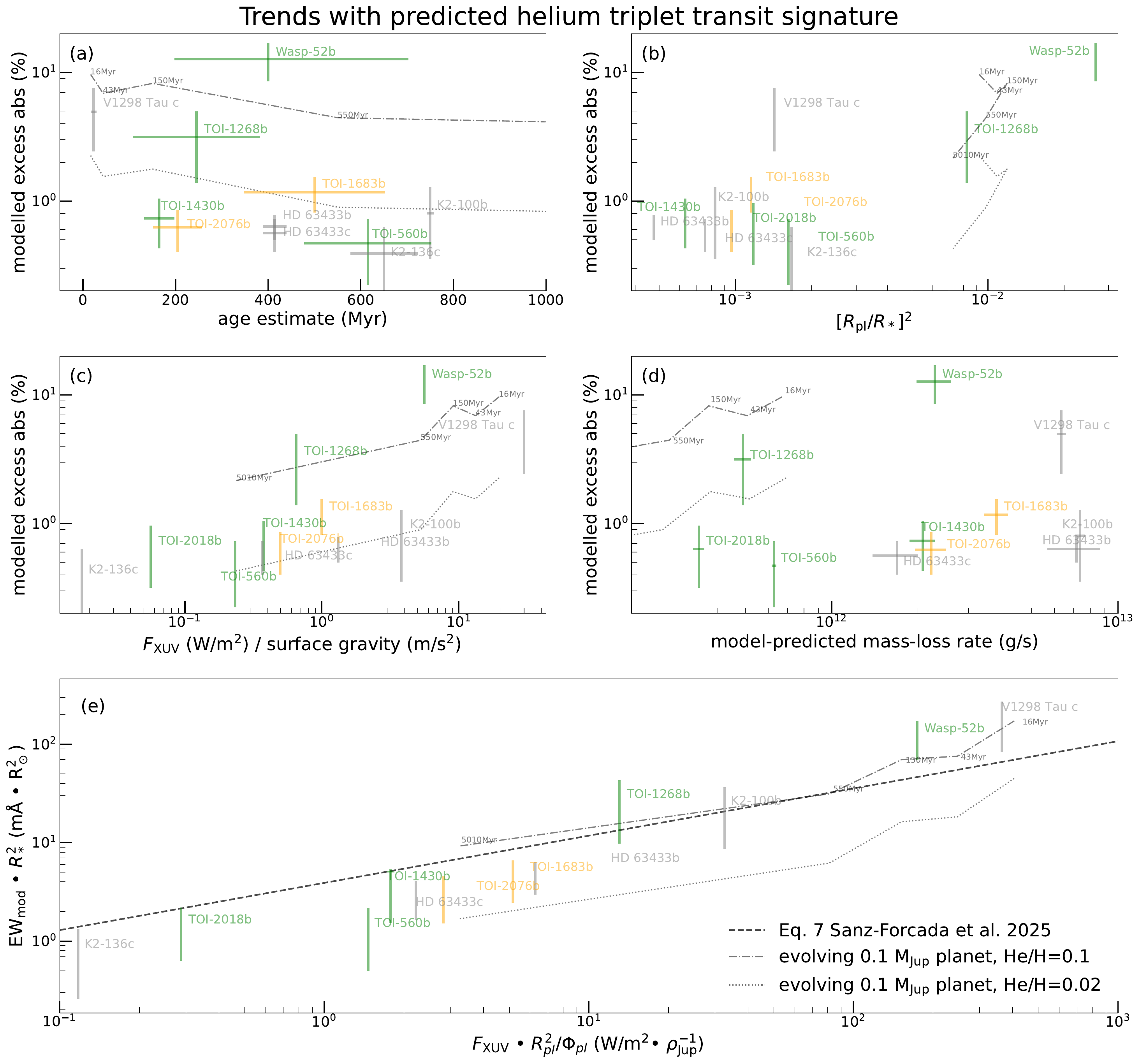}
    \caption{  The same as Figure \ref{fig:trends_obs}, now relating model-predicted rather than observed helium triplet parameters (y-axes) to various system parameters (x-axes). For each planet, the spread in the predicted excess absorption, the equivalent width EW$_{\text{mod}}$ and the mass-loss rate corresponds to assuming a He/H number fraction ranging from 0.02 to 0.1. The dashed-dotted and dotted tracks are that of theoretical, evolving, 0.1$~$M$_{\text{Jup}}$ planets with respective He/H number fractions of 0.1 and 0.02, orbiting an initially fast rotating K-dwarf at 0.045$~$au \citep{Alam_2024_young_non_detects, Allan_et_al_He_evol_2024}. Colour again indicates each planet's observational outcome, with green for detections, amber for conflicting interpretations (see Table \ref{Tab:observational}), and grey for non-detections. }
 \label{fig:trends_modelled}
\end{figure*}

Many studies have sought relations between various planetary parameters and the helium triplet signature \citep{Allart_2023_SPIRou_He_11_gas_giants, Orell-Miquel_2024_MOPYS, Krishnamurthy_Cowan_2024, Linssen_2024B_sunSET,McCreery_2025, Ballabio_Owen_2025}. We now add to this discussion the insights gained from our work here. We do so by discussing relations (or lack thereof) between planetary parameters and the observed absorption (Figure \ref{fig:trends_obs}), as well as the model-predicted absorptions (Figure \ref{fig:trends_modelled}).

From their homogeneous sample of eleven observations, \citet{Allart_2023_SPIRou_He_11_gas_giants} noted a trend between stellar age and triplet absorption, but did not draw further conclusions on this given the small sample size and lack of precision in the age estimates. \citet{Orell-Miquel_2024_MOPYS} instead found no clear trend with age in their larger survey, in agreement with the findings of \citet{Krishnamurthy_Cowan_2024}, as mentioned in section \ref{sec:intro}. This lack of a clear trend is again shown in Figure \ref{fig:trends_obs}a. On first glance, this may seemingly contradict our previous theoretical works, which modelled the atmospheric escape over the entire evolution of individual planets \citep{Allan_Vidotto_2019,Allan_et_al_He_evol_2024}. These theoretical studies both demonstrated a clear trend with younger ages and deeper absorptions. 
However, applying the same self-consistent hydrodynamic modelling to the diverse sample of observed young planets, it is clear that the resulting model predictions shown in Figure \ref{fig:trends_modelled}a, also exhibit no strong trend between age and absorption. 
This lack of trend with age is therefore in agreement with the current observations. Hence, for an individual planet along its evolution, its atmospheric escape and corresponding detectable signature are greatest at younger ages, as shown by the tracks of theoretical evolving 0.1$~$M$_{\text{Jup}}$ planets orbiting an initially fast rotating K-dwarf at 0.045$~$au \citep{Alam_2024_young_non_detects, Allan_et_al_He_evol_2024} in figure \ref{fig:trends_modelled}. However, for a diverse sample of exoplanets, both models and observations are in agreement that differences in other system parameters more strongly affect the resulting triplet signature, compared to the effect of system age. Hence, studies seeking to observe atmospheric escape should place such parameters above age in their consideration of potential targets.   %

One such property is the geometric transit depth given by $(R_{\text{pl}}/R_*)^2$. Figures \ref{fig:trends_obs}b and \ref{fig:trends_modelled}b show the relation between the geometric transit depth and the observed and predicted helium triplet excess absorptions, respectively. For \textit{both} observed and model signatures, a relation is evident amongst the planets with detections. In each case, the strong detections of the hot-Saturns Wasp-52b and TOI-1268b are largely driving this trend. The planets with non-detections being to the lower end of the geometric transit depth scale, further strengthens this noted trend. Similarly, \citet{Krishnamurthy_Cowan_2024} note a preference for larger planetary radii amongst successful detections in their reanalysis of published studies. They suggest that the various non-detections for older systems with smaller planets are not surprising due to an insufficient accumulation of primordial atmosphere and their vulnerability to losing these atmospheres.

Another parameter which shows a positive trend with both the observed and modelled helium triplet signatures is the ratio of XUV flux, $F_{\text{xuv}}$ to the planetary gravity (a popular proxy for the mass-loss rate). To obtain an $F_{\text{xuv}}$ for each planet, we sum the individual X-ray, hEUV and sEUV fluxes, which depend on the respective stellar luminosities of these wavelength bins and the planet's orbital distance. In both the observed (figure \ref{fig:trends_obs}c) and modelled (figure \ref{fig:trends_modelled}c) triplet signatures of planets with successful detections (coloured in green), a positive relation is evident. A large value for this ratio alone however is a poor indicator of a strong detection compared to a large geometric transit depth, as shown by the many non-detections (in grey) for planets with large XUV flux to gravity ratios. Furthermore, it is reliant upon the received XUV flux, a parameter which is greatly unknown for the majority of systems.

Interestingly, the systems with the strongest atmospheric escape rates as predicted by our hydrodynamic model have yielded non-detections (V$~1298~$Tau$~$c, K2-100b, HD$~$63433b) as seen in panel \ref{fig:trends_obs}d. As discussed in section \ref{sec:triplet_predictions}, these planet's host stars have the strongest X-ray surface fluxes of our sample, which implies that they also have stronger stellar winds that could potentially confine the extension of the helium-absorbing planetary atmospheres.
Evidently, we do not find clear a trend between the predicted mass-loss rates and the predicted (figure \ref{fig:trends_modelled}d), nor the observed (figure \ref{fig:trends_obs}d) helium absorptions. \citet{Ballabio_Owen_2025} recently demonstrated that the helium triplet excess absorption scales linearly with mass-loss rate, with all other things being equal. As many system properties vary substantially for the exoplanets of our sample (see tables \ref{Tab:pl_inputs} and \ref{Tab:stellar_SED}), the lack of trend is to be expected. \citet{Ballabio_Owen_2025} also show that absorption is weaker for hotter atmospheric outflows. The high temperatures of K2-100b and HD$~$63433b with respective peaks of 11.3 and 10.3kK, could also have contributed to their predicted weak signatures, consistent with their tightly constrained non-detections. 
However, we also find that planets with even hotter peak temperatures and weaker escape rates such as Wasp-52b are capable of producing strong absorptions, highlighting the importance of additional system parameters in affecting the resulting helium triplet signature. 

In short, we find that a strong detection does not necessarily imply a strong rate of atmospheric escape nor does a weak or non-detection imply a weak escape rate. This point is better illustrated by the tracks of the evolving, theoretical 0.1$~$M$_{\text{Jup}}$ planets in figure \ref{fig:trends_modelled}. Despite the stronger mass-loss rate at the younger age of 43$~$Myr, the excess absorption is predicted to be stronger at 150$~$Myr (figure \ref{fig:trends_modelled}d). As explained previously in \citet{Allan_et_al_He_evol_2024}, this is due to a larger ratio of planetary to stellar radii inputs adopted at the age of 150$~$Myr (figure \ref{fig:trends_modelled}b). While both planetary and stellar radii shrink substantially during the system's early evolution, the more rapid decay of the stellar radius in this case results in a stronger helium detection despite a weaker escape at 150$~$Myr. Therefore, consideration of the stellar radius could be of benefit when selecting candidates for future helium detections, in addition to the more obvious XUV flux and planetary gravity.

\citet[][section 5.3]{Zhang_2022_TOI_560} relates an order-of-magnitude empirical estimate for planetary mass-loss rate to EW$_{\text{obs}} \cdot R_{*}$, where EW$_{\text{obs}}$ is the equivalent width of the observed helium triplet profile. They do so by estimating the mass of obscuring helium triplet material. \citet{Zhang_2023_He_mature} then demonstrated a positive relation between EW$_{\text{obs}} \cdot R_{*}$ and $F_{\text{xuv}}  / \rho_{\text{xuv}}$ for a sample of helium observed exoplanets, the latter term arising from the energy-limited mass-loss approximation, with $\rho_{\text{xuv}}$ being the average planetary density calculated using the XUV photosphere radius. \citet[][]{Sanz-Forcada_2025_HeI_10830} noted the similar, strong, empirical relation: 
\begin{equation}
    EW_{\text{obs}} \cdot R_{*}^{2} \propto F_{\text{xuv}} ~ R_{\text{pl}}^2 / \Phi_{g},
\end{equation}
where $\Phi_g=G M_{\text{pl}}/R_{\text{pl}}$ is the planet gravitational potential with $G$ being the gravitational constant, and $R_{\text{pl}}^2 / \Phi_{g}$ is equivalent to the inverse average planet density. Figure \ref{fig:trends_obs}e shows this relation for our considered sample of exoplanets.

Figure \ref{fig:trends_modelled}e shows the same trend, only now replacing $EW_{\text{obs}}$ with our model-predicted equivalent width, EW$_{\text{mod}}$. 
In both cases, a clear trend is evident amongst the confirmed (green) and conflicted (yellow) detections.  
The theoretical individual planets are also seen to follow the trend of \citet[][]{Sanz-Forcada_2025_HeI_10830} reasonably well as they evolve over time. Their evolution tracks highlight the importance of the size of the stellar disk in the y-axis term. The previously discussed observational enhancement at 150$~$Myr is descaled through multiplication with the stellar disk area, causing the y-axis values at the sampled ages to decrease chronologically, more consistent with their decreasing atmospheric escape. 

Interestingly, the four planets with tightly constrained upper limits fall below the dotted line of the observational relation in Figure \ref{fig:trends_obs}e, while their model predictions yield values closer to the general trend in Figure \ref{fig:trends_modelled}e. This could indicate an overestimated XUV flux, an underestimated planetary gravity, or the omission of important physical processes such as stellar wind interactions from the model, the latter being supported by the high surface X-ray fluxes of their host stars, indicative of strong stellar winds (section \ref{sec:triplet_predictions}). The non-simultaneity of the helium triplet observations and the stellar X-ray observations, as well as the EUV flux relying mostly on extrapolations from the X-ray, complicate the selection of the XUV flux. This important parameter strongly affects both the hydrodynamic escape and its resulting observables. 
On the possibility of underestimating the planetary gravities, three of these four planets have planetary masses obtained via mass-radius relation (see Table \ref{Tab:pl_inputs}). Alternatively, the lower limit assumption of He/H=0.02 could also overestimate the true atmospheric helium abundance of these particular planets. %

\section{Conclusions}
\label{sec:conclusions}
We applied our hydrodynamic model of atmospheric escape to a sample of young exoplanets which have helium triplet observations. We modelled their atmospheric escape as well as their corresponding helium triplet transit signature, achieving a good overall agreement between model and observation. From this work, we draw the following conclusions:

\begin{itemize}
    \item For an individual planet along its evolution, its atmospheric escape and corresponding detectable signature are greatest at younger ages. However for a diverse sample of exoplanets, differences in alternative parameters have a stronger effect on the resulting triplet signature, compared to the effect of age. This is seen both observationally and in our model predictions of the same systems.

    \item Our model predictions reproduce well the strong empirical trend of \citet{Sanz-Forcada_2025_HeI_10830}: $ EW \cdot R_{*}^{2} \propto F_{\text{xuv}} \cdot R_{\text{pl}}^2 / \Phi_{g}$. %

    \item Studies seeking to observe atmospheric escape should place parameters such as the geometric transit depth or \textbf{$F_{\text{XUV}} \cdot R_{\text{pl}}^2 / \Phi_{g}$} above age in their consideration of potential targets for future helium triplet detections.   
  
    \item Studies modelling the helium triplet transit signature should pay careful consideration to their choice of assumed lower boundary conditions. This is particularly the case when modelling small, mini-Neptune size planets for which the common assumption of setting the base of the modelled upper atmosphere to $1~R_{\text{pl}}$ can introduce large uncertainty in the predicted triplet profile.

    \item Our model predictions favour a higher He/H number fraction for smaller mini-Neptunes planets compared to hot-Saturns. This can be interpreted physically by the mini-Neptunes undergoing atmospheric escape with a preferential loss of hydrogen over helium in combination with a sufficiently small reservoir of atmospheric material so that the previously escaped material can affect the current He/H ratio. Further observations and modelling of more of each class of planet would be needed to confirm such behaviour. 

\end{itemize}

As the number of observations continues to grow, the known trends between various planetary parameters and the observed helium triplet signature of escape will become increasingly reliable, while new trends will no doubt be revealed. These trends can improve our approach of target selection for future observations.

\section*{Acknowledgements}
We thank the anonymous referee for their insightful comments which have greatly improved this work. APA and AAV acknowledge funding from the European Research Council (ERC) under the European Union's Horizon 2020 research and innovation programme (grant agreement No 817540, ASTROFLOW). We thank several colleagues for helpful discussions over this project, particularly regarding the lower boundary conditions: Dr.~Luca Fossati, Dr.~James Owen, Dr.~Carolina Villarreal D’Angelo.

\section*{Data Availability}
The data described in this article will be shared on reasonable request to the corresponding author.

\bibliographystyle{mnras}
\bibliography{example} %

\bsp	%

\appendix

\section{Sensitivity to the assumed density at the base of the atmosphere}

\begin{figure}
    \includegraphics[width=0.47\textwidth]{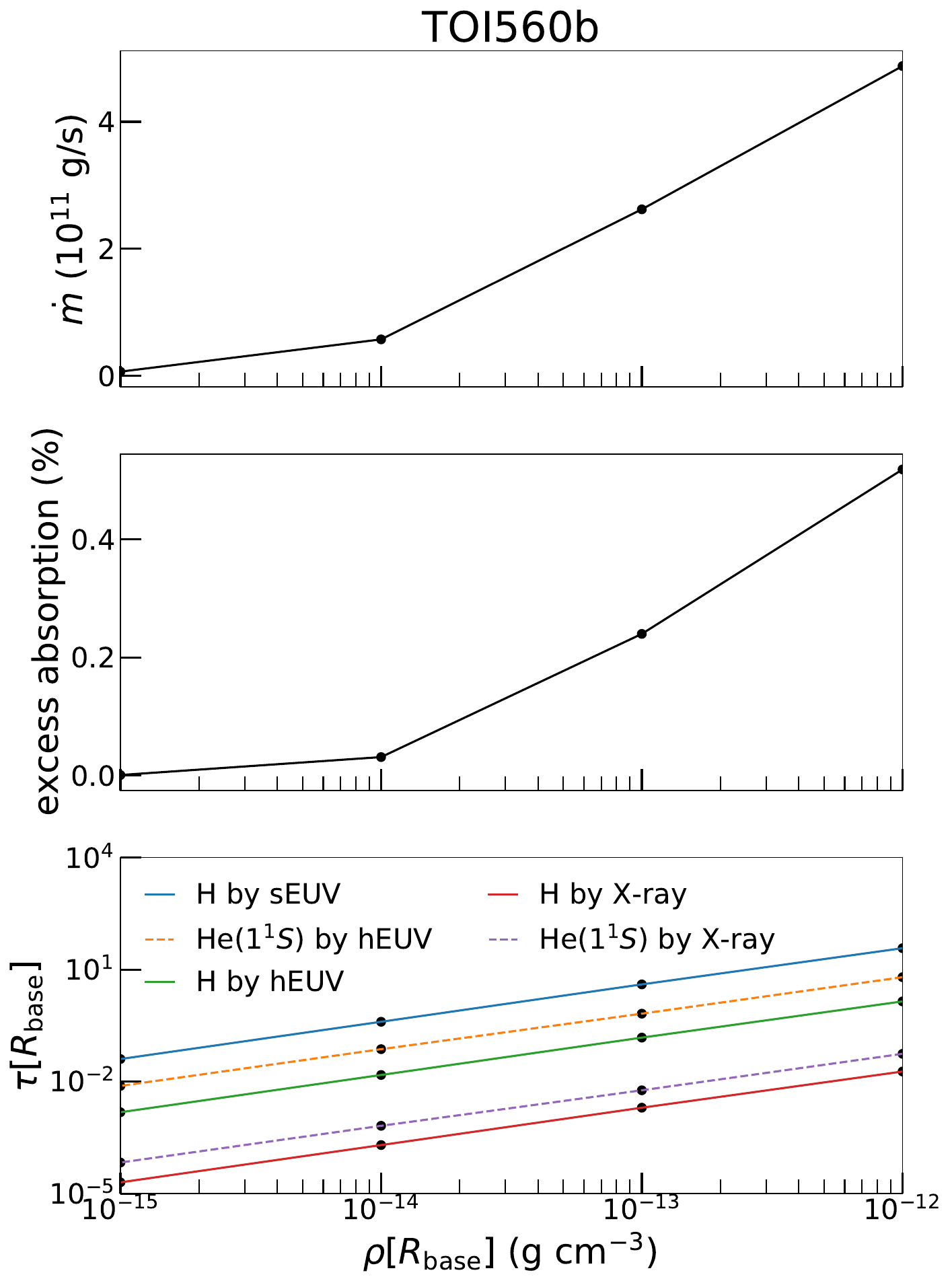}
    \caption{ Sensitivity of the mass-loss rate (upper-panel) and the the phase averaged helium 1083nm excess absorption (central-panel) to the assumed density at the base of the modelled atmosphere, while holding $R_\text{base} = 1~R_\text{pl}$ and $T[R_{\text{base}}]=1000~$K constant. The remaining planetary parameters are that of TOI-560b, and assuming He/H=0.1. The lower panel displays optical depths to various photoionisations. }
 \label{fig:sensitivity_rho0_560b}
\end{figure}

\begin{figure}
    \includegraphics[width=0.47\textwidth]{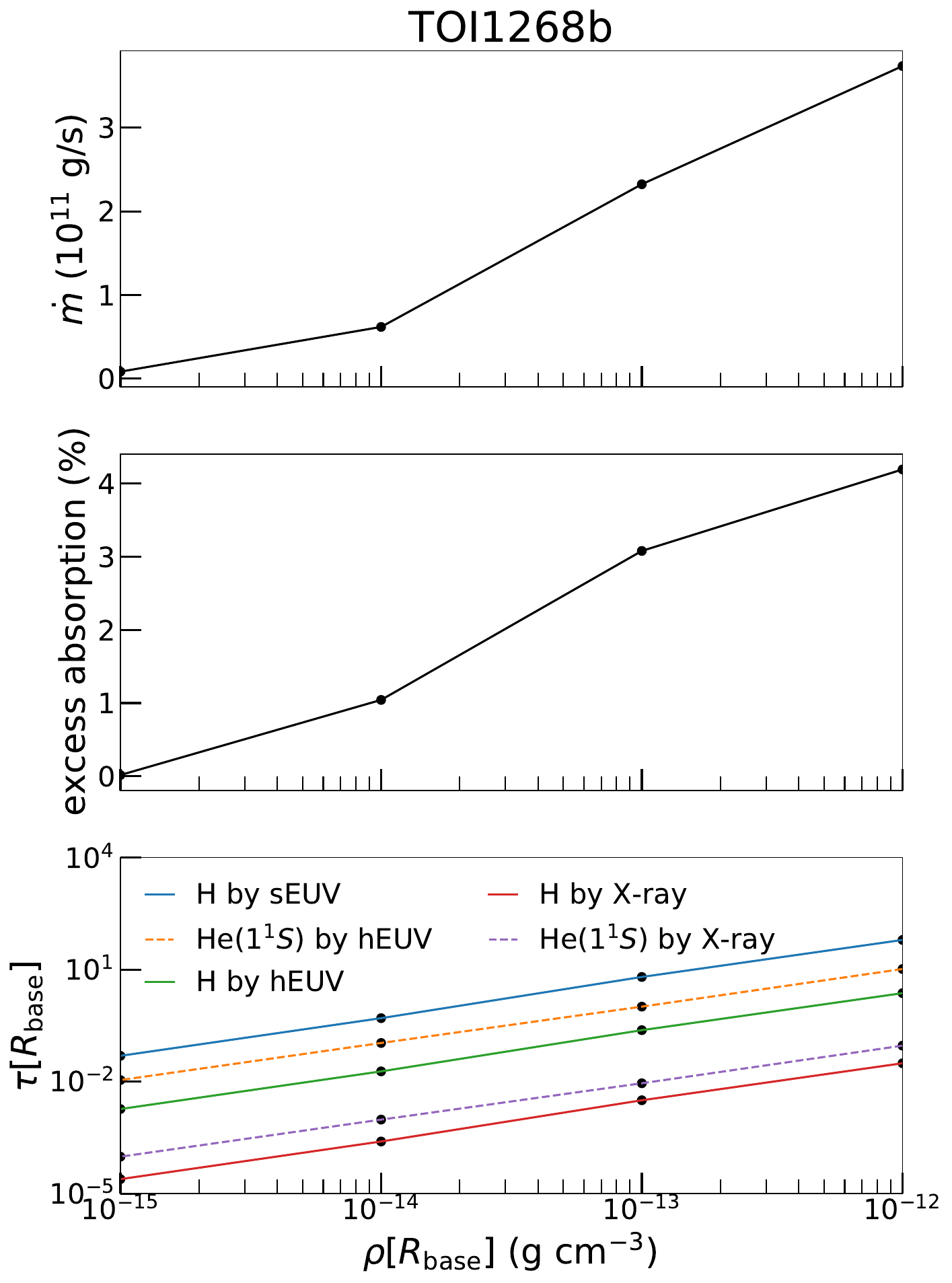}
    \caption{ The same as figure \ref{fig:sensitivity_rho0_560b} again assuming $R_\text{base} = 1~R_\text{pl}$ and $T[R_{\text{base}}]=1000~$K and varying $\rho[R_\text{base}]$, while the remaining planetary parameters are now that of TOI-1268b.}
 \label{fig:sensitivity_rho0_1268b}
\end{figure}

In order to emphasize the need for its careful selection, Figures \ref{fig:sensitivity_rho0_560b} and \ref{fig:sensitivity_rho0_1268b} demonstrate the sensitivity of our model predicted atmospheric escape and its corresponding helium triplet signature to the density assumed at the base of the planetary atmosphere. Here, our calculations start at $R_\text{base} = 1~R_\text{pl}$ and we use $T[R_{\text{base}}]=1000~$K.  
As discussed in section \ref{sec:determine_atm_base_low_BC}, our model exhibits a strong sensitivity of the predicted helium triplet profile with the assumed base density, beyond the uncertainties of the helium triplet observations. 
As shown in the appendix of \citet{Murray-Clay2009} and confirmed by our \citet{Allan_Vidotto_2019} model, for modelling considering photons concentrated at a single energy bin (20 eV) and with a single possible absorber (neutral hydrogen), the predicted escape properties plateaus to stable values for various assumed densities at the atmospheric base, once a sufficiently high density is reached, and an optical depth for photoionisations at the base of the model is above unity. The number of photons that penetrate the atmosphere is related to photoionisation and heating, which drives the planetary escape. However, with the inclusion of more photon energies and potential absorbers, this plateau behaviour does not occur at a reasonable base density (see top panels of Figures \ref{fig:sensitivity_rho0_560b} and \ref{fig:sensitivity_rho0_1268b}). This is because as we raise the base density, only some optical depths of possible photoionisations reach above unity, which implies that raising the base density increases the number of one of the possible photoionisations (affecting heating deposition at the lower part of the atmosphere)  as is seen in the lower panels of figures \ref{fig:sensitivity_rho0_560b} and \ref{fig:sensitivity_rho0_1268b}, rather than a saturation of photoionisations being reached. Across a wide range of assumed base densities or pressures, the model remains sensitive to the choice of base density due to the number of occurrences of at least one of the potential photoionisations increasing.

\section{Models assuming commonly used lower boundary conditions}
\label{sec:appendix_old_BC} 

\begin{figure*}
    \includegraphics[width=0.95\textwidth]{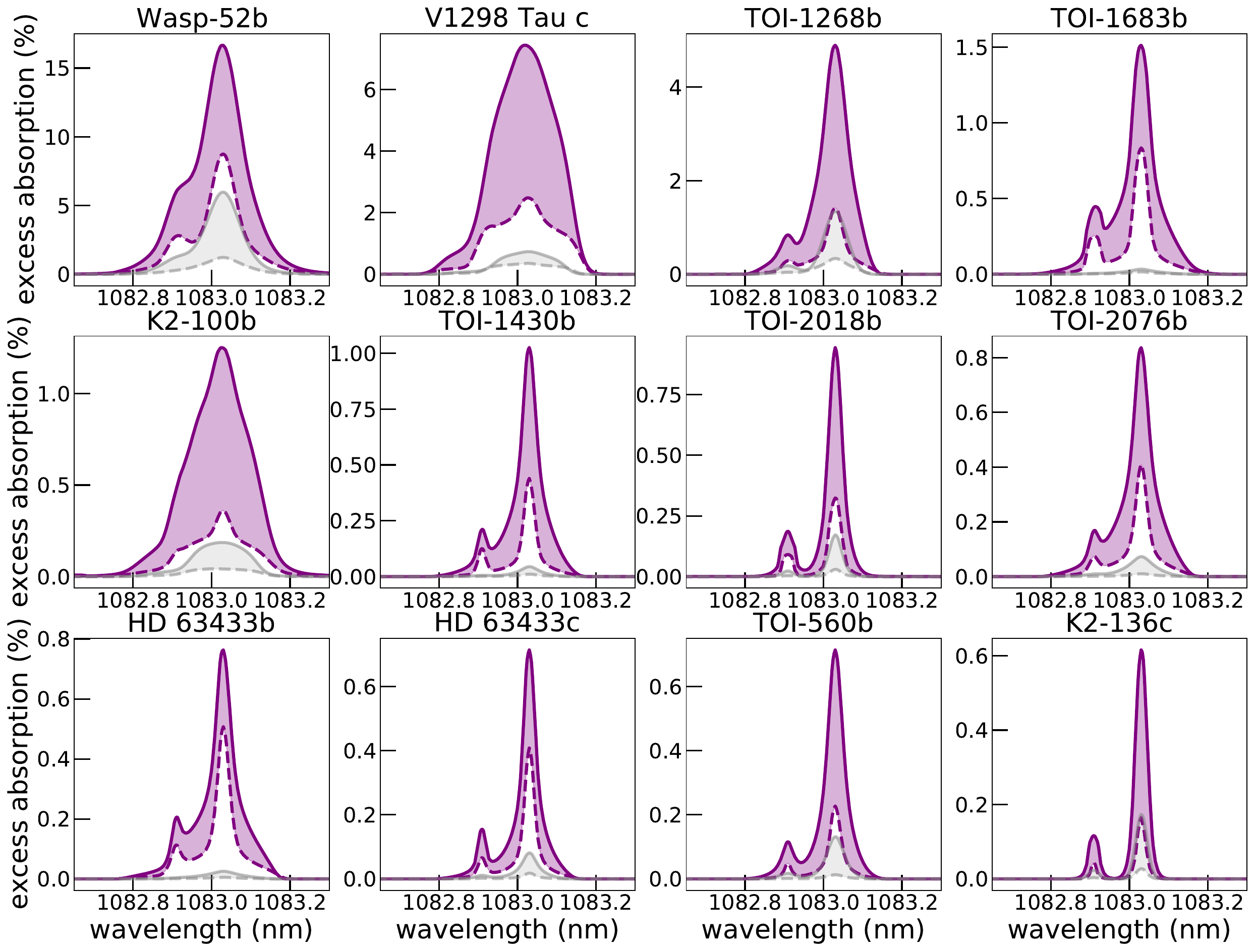}
    \caption{ Comparison between the predicted triplet profiles with our calculated lower boundary conditions (purple, same as that in figure \ref{fig:all_helium_profiles_10nanobar}) and those predicted by reverting to the standardly assumed lower boundary conditions of $R_\text{base}=1~R_\text{pl}$, $\rho[R_{\text{base}}]=4 \times 10^{-14}$g$~$cm$^{-3}$, $T[R_{\text{base}}]=1000~$K (grey).  }
 \label{fig:all_helium_profiles_compared_to_old_assumptions}
\end{figure*}

In section \ref{sec:determine_atm_base_low_BC} we outline our reasoning for choosing more physically motivated properties at the lower boundary of our modelled atmospheres. In section \ref{sec:triplet_predictions} we showed our predicted triplet signatures with our improved lower boundary assumptions. For comparison, in figure \ref{fig:all_helium_profiles_compared_to_old_assumptions} we now show the model's predicted triplet signatures assuming the lower boundary conditions assumed in \citet{Allan_et_al_He_evol_2024} and more akin to that used in the current literature, $R_\text{base}=1~R_\text{pl}$, $\rho[R_{\text{base}}]=4 \times 10^{-14}$g$~$cm$^{-3}$, $T[R_{\text{base}}]=1000~$K.

\section{Observational properties}
\begin{table*}
\caption{Observational summary of the considered sample of exoplanets.}
\label{Tab:observational}
\begin{tabular}{llllll}
\toprule
name & interpretation & peak abs (\%) & EW (m\AA\ ) & telescope & Reference \\ \midrule
Wasp-52b & detect & $3.44\pm{0.31}$ & $39.583\pm{1.4}$ & Keck II/NIRSPEC & \citep{Kirk_2022_Wasp52b_Wasp177b} \\
V1298$~$Tau$~$c & upper limit & $<1.13$ & $<8.49$ & Keck II/NIRSPEC & \citep{Alam_2024_young_non_detects} \\
V1298$~$Tau$~$c & upper limit & $<3.75$ & $<95.84$ & CARMENES & \citep{Orell-Miquel_2024_MOPYS} \\
TOI-1268b & detect & $2\pm{0.16}$ & $19.1\pm{1.8}$ & CARMENES & \citep{Orell-Miquel_2024_MOPYS} \\
TOI-1683b & detect & $0.84\pm{0.17}$ & $8.5\pm{1.6}$ & Keck/NIRSPEC & \citep{Zhang_4_mini_Nep} \\
TOI-1683b & upper limit & $<0.7$ & - & CARMENES & \citep{Orell-Miquel_2024_MOPYS} \\
K2-100b & upper limit & $<0.56$ & $<4.44$ & Keck II/NIRSPEC & \citep{Alam_2024_young_non_detects} \\
K2-100b & upper limit & $<1.2$ & $<5.7$ & IRD/Subaru & \citep{Gaidos_2020_K2100b} \\
TOI-1430b & detect & $0.91\pm{0.11}$ & $9.5\pm{1.1}$ & CARMENES & \citep{Orell_Miguel_650_Myr_old_sub_Nep_HD235088b} \\
TOI-1430b & detect & $0.64\pm{0.06}$ & $6.6\pm{0.5}$ & Keck/NIRSPEC & \citep{Zhang_4_mini_Nep} \\
TOI-2018b & detect & $1.02_{-0.22}^{+0.19}$ & $7.8\pm{1.5}$ & CARMENES & \citep{Orell-Miquel_2024_MOPYS} \\
TOI-2076b & detect & $1.01\pm{0.05}$ & $10.0\pm{0.7}$ & Keck/NIRSPEC & \citep{Zhang_4_mini_Nep} \\
TOI-2076b & upper limit & $<1.0$ & - & CARMENES & \citep{Orell-Miquel_2024_MOPYS} \\
HD$~$63433b & upper limit & $<0.47$ & $<2.52$ & Keck II/NIRSPEC & \citep{Alam_2024_young_non_detects} \\
HD$~$63433b & upper limit & $<0.5$ & - & Keck II/NIRSPEC & \citep{Zhang_2022_HD63433_upperlimit_non_detect} \\
HD$~$63433b & upper limit & $<0.34$ & $<2.0$ & CARMENES & \citep{Orell-Miquel_2024_MOPYS} \\
HD$~$63433c & upper limit & $<0.4$ & $<4.2$ & CARMENES & \citep{Orell-Miquel_2024_MOPYS} \\
HD$~$63433c & upper limit & $<0.5$ & - & Keck/NIRSPEC & \citep{Zhang_2022_HD63433_upperlimit_non_detect} \\
TOI-560b & detect & $0.72\pm{0.08}$ & $8.6\pm{0.6}$ & Keck/NIRSPEC & \citep{Zhang_4_mini_Nep} \\
TOI-560b & detect & $0.64\pm{0.08}$ & $7\pm{0.4}$ & Keck/NIRSPEC & \citep{Zhang_2022_TOI_560} \\
K2-136c & upper limit & $<2.3$ & $<25$ & Subaru/IRD & \citep{Gaidos_2021_He_non_detect_K2136c} \\
\bottomrule \end{tabular}
\end{table*}

Table \ref{Tab:observational} compiles observational data on all planets considered in this work. We consider only high-resolution observations. In the figures of relations of properties with the helium triplet signature in section \ref{sec: trends}, we choose only one representative observation for each planet. This is either the most recent, highest resolution detection, or most tightly constrained non-detection. The chosen observation for each planet is the first row to appear for each planet in table \ref{Tab:observational}. Due to conflicting findings for the observations of TOI-2076b and TOI-1683b, we mark them in amber in figures \ref{fig:trends_obs} and \ref{fig:trends_modelled}.

\section{Comparison of predicted mass-loss rates}
\label{sec:compar_predict_mdot_McCreery}

\begin{table}
    \caption{Ratios of the mass-loss rates reported in \citet{McCreery_2025} relative to that predicted here, assuming both a high and low He/H fraction. $^*$While we did not run He/H=0.01 models, we consider our He/H=0.02 models sufficient for this comparison, as the assumed fraction only minorly affects our hydrodynamic predictions (see Table \ref{Tab:escape_predictions}), compared to the strong dependence with the mass-loss rates of  \citet{McCreery_2025}, as explained in the text. }
    \begin{tabular}{ccc}
        \toprule
        planet & ratio assuming He/H=0.1  & ratio assuming He/H=0.01$^*$ \\
         & (\%) & (\%) \\
        \midrule
        Wasp-52b & 6--12  & 15--29  \\
        TOI-560b & 3.3--22  & 10--66  \\
        TOI-1430b & 0.08--0.48 & 0.4--1.9  \\
        TOI-1683b & 0.03--0.63 & 0.2--4.2  \\
        TOI-2076b & 0.08--0.2 & 0.2--0.4  \\
        \bottomrule
    \end{tabular}
    \label{tab:mdot_compar_McCreery}
\end{table}

It was mentioned in section \ref{sec:atmo_escape_predictions} that our mass-loss rate predictions are strong relative to some recent studies of the same planets \citep{Zhang_4_mini_Nep, McCreery_2025}. %
The latter of these studies modelled various helium triplet observed exoplanets, including five planets common to our sample using the atmospheric escape code, \textsc{p-winds} \citep[][version 1.4.6]{pwinds, dos_santos_2023_10419485}. Table \ref{tab:mdot_compar_McCreery} shows that the mass-loss rates predicted here are considerably stronger than that found in \citet{McCreery_2025}. Our predictions also differ in that their mass-loss rate predictions are strongly dependent on the assumed He/H fraction, whereas our predictions are not. This is due to their  mass-loss predictions involving modelling of the planets' helium triplet profiles, which are strongly affected by the assumed abundance of helium. Whereas, our mass-loss rates are predicted purely by our hydrodynamic model and hence are only weakly affected by the assumed helium abundance as mentioned in section \ref{sec:atmo_escape_predictions}. 
In general, the different nature of the approaches taken in our studies makes it difficult to understand the reasons behind the noted mass-loss rate differences. For example, our model self-consistently outputs a mass-loss rate and atmospheric profiles of temperature and velocity, whereas the mass-loss rate and outflow temperature are free parameters in the set-up of \citet{McCreery_2025}. Both the temperature and the mass-loss rate (or base density) strongly affect the velocity and density profiles in models like theirs, which assume a constant temperature (or sound speed) throughout the atmospheric structure. Their approach however benefits from utilising helium triplet observations to retrieve good fits to the observed data, whereas, our purely forward-model predictions might not.

\label{lastpage}
\end{document}